\documentclass[twocolumn,pra,aps,showpacs]{revtex4}
\usepackage{graphicx}
\usepackage{amsmath}
\usepackage{amssymb}

\begin{document}

\title{Concatenated beam splitters, optical feed-forward and the nonlinear sign gate}

\author{Kurt Jacobs}

\affiliation{Quantum Science and Technologies Group, 
Hearne Institute for Theoretical Physics, Department of Physics and Astronomy, 
202 Nicholson Hall, Tower Drive, Baton Rouge, LA 70803, USA}

\author{Jonathan P. Dowling}

\affiliation{Quantum Science and Technologies Group, 
Hearne Institute for Theoretical Physics, Department of Physics and Astronomy, 
202 Nicholson Hall, Tower Drive, Baton Rouge, LA 70803, USA}

\begin{abstract}
We consider a nonlinear sign gate implemented using a sequence of two
beam splitters, and consider the use of further sequences of beam
splitters to implement feed-forward so as to correct an error
resulting from the first beam splitter. We obtain similar results to
Scheel {\em et al.} [Scheel {\em et al.}, Phys. Rev. A {\bf 73},
034301 (2006)], in that we also find that our feed-forward procedure
is only able to produce a very minor improvement in the success
probability of the original gate.
\end{abstract}


\maketitle

The use of measurement, coupled with linear optical elements can
produce, albeit probabilistically, effective optical nonlinearities, a
fact which was first realized in the paper of Knill, Laflamme and 
Milburn~\cite{KLM}. This is of considerable
interest because, as demonstrated in this work, such nonlinearities
can be used to construct quantum gates, which, even though they
succeed with a less that unity probability can, at least in principle,
be used for reliable quantum computing~\cite{Kok06}. An example of
such a gate is the nonlinear sign gate, which transforms an input
state $\alpha|0\rangle + \beta|1\rangle + \gamma|2\rangle$ by flipping
the sign of $\gamma$~\cite{KLM}.

Such optical gates are implemented by sending a single optical mode
successively through a sequence of linear optical elements, where the
mode may interact with auxiliary modes at one or more beam splitters.
The auxiliary modes are measured (using photocounters), and if the
measurement results are right, the correct nonlinear transformation
will be implemented on the input mode. While in most of these schemes
the optical elements are independent of the measurement results, it is
possible to chose successive optical elements based on the results of
measurements performed on modes that have interacted with the input
mode in preceding elements. Such a procedure is an example of
feedback~\cite{FBRev}, which in this case is referred to as {\em feed-
forward}, because the change which is implemented as a result of the
measurement is implemented ``downstream'' at a later point in the
sequence.

Here we investigate the use of feed-forward in the implementation of
the nonlinear sign (NS) gate. In the absence of feed-forward, it has
been shown that the maximum probability with which the gate can be
implemented is 1/4~\cite{Scheel04,Eisert05}, and further that 1/2 is an
absolute upper bound~\cite{Knill03}. The question is therefore whether
feed-forward can be used to significantly increase the probability
above 1/4. An investigation of this question has already been made by
Scheel {\em et al.}~\cite{Scheel05}. Their approach was first to implement the gate,
and if the measurement results were incorrect, to feed the output into
a further linear optical measurement sequence to correct the
transformation. Here we consider an alternative implementation of the
NS gate which breaks down into two sequential steps, each involving a
beam splitter and a measurement on the associated auxiliary
mode~\cite{Ralph02,Rudolph01}. This allows us to modify the elements
halfway through the operation of the gate if the measurement at the
first beam splitter was not the required one.
 
\begin{figure}
\includegraphics[width=7.5cm]{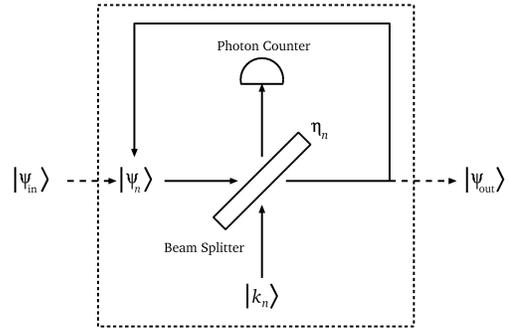} 
\caption[example]{
An optical ``feedback'' circuit showing one method of implementing the
``feed-forward'' circuits investigated in the text. In this
configuration, the beam is repeatedly sent through a beam splitter
whose transmitivity is adjusted at each pass, until the required
transformation of the input is obtained at the output mode.}
\label{fig1}
\end{figure}

Since a sequence of two beam splitters implements a transformation
containing two free parameters, one can expect to be able to perform a
correction using this configuration. Here we will restrict ourselves to
correction circuits of this form. The resulting complete gate,
including feed-forward, will therefore consist of a sequence of
concatenated beam splitters, where the number used depends upon the
sequence of measurement results. We note that such a configuration
could alternatively be implemented with a single controllable beam
splitter: when the beam passes through the beam splitter a measurement
is made on the auxiliary mode. Depending on the result, the beam is
either routed back through the beam splitter for further
transformation, or output if the transformation has been
completed. This feedback procedure is illustrated in
figure~\ref{fig1}.

Scheel {\em et al.} found that they were only able to obtain a very 
minor improvement in the overall success probability of the gate, and 
concluded that feed-forward did not appear likely to be a useful tool 
in the generation of optical nonlinearities. We also report here a 
negative result; we find that the feed-forward procedure described above 
is only able to achieve a small increase in the success probability 
of the initial two-stage gate from which it builds. 

\begin{figure}
   \includegraphics[width=8cm]{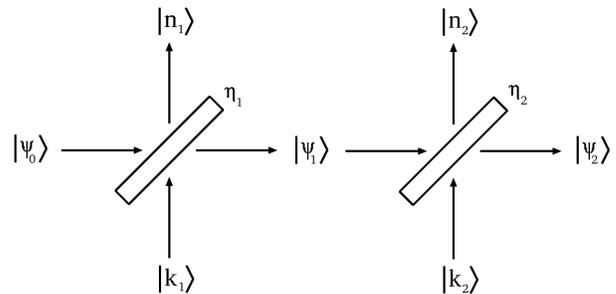} 
\caption[example]{ 
A sequence of two beam splitters, each accompanied by a measurement on
their auxiliary output modes. Such a configuration performs a
nonlinear transformation on the input state $|\psi\rangle$ with some
probability P. The NS gate of Ralph et al. and Rudolph and Pan has $k_1=n_1=1$, $k_2=n_2=0$ and a success probability of $P=0.2265$. The NS gate can also be implemented
with $k_1=n_1=k_2=n_2=1$, giving a success probability of $P=0.209$.}
\label{fig2}
\end{figure}

To begin let us describe the operation of the two-stage gate devised
by Ralph {\em et al.}~\cite{Ralph02} and Rudolph and 
Pan~\cite{Rudolph01} (Note that
while we consider standard beam splitter in the following analysis,
this gate can also be implemented using polarization beam splitters,
which are likely to be easier in practice).  This involves mixing the
input beam with an auxiliary mode at a single beam splitter,
measuring the auxiliary mode, and then repeating the process at a
second beam splitter with a second auxiliary mode. This configuration
is depicted in Figure~\ref{fig2}. One injects a single photon state
into the first auxiliary mode, and a vacuum state into the second.
The correct transformation is obtained if the input photon is detected
in the first auxiliary mode output, and zero photons are detected in
the second output. The reason that this works is because each beam
splitter and measurement combination performs a nonlinear
transformation on the input state. If we send in $k$ photons in the
auxiliary mode of the beam splitter, and detect all of them at the
auxiliary output, then no photons have been added to the input
mode. In this case, the coefficients of the input state transform as
\begin{eqnarray}
  \alpha' & = & \alpha \sqrt{\eta}^k \\  
  \beta'  & = & \beta \sqrt{\eta}^{k+1} \left[ 1 - k \xi \right] \\ 
  \gamma' & = & \gamma \sqrt{\eta}^{k+2} \left[ 1 - 2k \xi
                 + k (k-1) \xi^2 \right]  ,
  \label{tran1}
\end{eqnarray}
where $\eta \in [0,1]$ is the beam splitter transmitivity, and we have 
defined $\xi = (1-\eta)/\eta$. The overall
norm given by ${\cal N} = |\alpha'|^2 + |\beta'|^2 + |\gamma'|^2$
gives the probability that this transformation occurs (i.e. the
probability that $k$ photons are detected at the output). If we fix
$k$, then their is one free parameter in the transformation, being the
beam splitter transmitivity $\eta$.

Note that since the output state must be normalized by dividing by
${\cal N}$, there are only two independent parameters in the state (so
long as the coefficients $\alpha$, $\beta$ and $\gamma$ remain 
real). Thus if we wish to perform any desired transformation on the
input state that preserves the reality of the coefficients, we need a
transformation that has at least two free parameters that can be
independently chosen. This is why two successive beam splitters are
required to implement the NS gate, since each beam slitter provides a
transformation with one free parameter. To perform the gate, we need
to chose both parameters, $\eta_1$ and $\eta_2$, so that the end 
result is merely a sign change for $\gamma$ (up to an overall scaling
factor). Note, however, that the fact we have two free parameters does
not guarantee we can find the transformation we need. For example, the
transformation obtained with $k=0$ re-weights the relative magnitudes
of $\alpha$, $\beta$, and $\gamma$, but only with positive factors. As
a result no number of beam splitters with $k=0$ can generate the NS
gate. This is why the gate requires one of the beam splitters to have
$k=1$.

The set of transformations displayed in Eqs.(\ref{tran1}) does not,
however, exhaust those which are possible with a beam splitter
followed by a measurement on the auxiliary output. Such an element
will also add or subtract $n$ photons from the mode if we detect $n$
fewer or $n$ more photons at the auxiliary output than we injected
into the auxiliary input. Doing so also generates a new set of
transformations, as well as changing the photon number. If we inject
$k$ photons, and detect $k-1$ photons, then for $k\geq 2$ the 
transformation becomes
\begin{eqnarray}
 \alpha' & = & \alpha \sqrt{k(k-1)(1-\eta)} \sqrt{\eta}^{k-1} \\  
 \beta'  & = & \beta \sqrt{2k(k-1)(1-\eta)}\sqrt{\eta}^k 
                     \left[ 1 - (k-1) \xi/2 \right] \\  
 \gamma' & = & \gamma \sqrt{(3/2)k(k-1)(1-\eta)} \sqrt{\eta}^{k+1} \nonumber \\
         &  &  \;\;\; \times  \left[ 1 - (k-1) \xi
                      + (k-1)(k-2) \xi^2/3 \right] ,
  \label{tran2}
\end{eqnarray}
where we have removed an overall phase factor of $-i$, and where the
unnormalized output state is now $\alpha' |1\rangle + \beta' |2\rangle
+ \gamma' |3\rangle$, since a photon has been added to the mode.  For
$k=1$ the transformation is $\alpha' = \alpha \sqrt{1-\eta}$, $\beta'
= \beta\sqrt{2\eta(1-\eta)}$ and $\gamma' = \gamma \eta
\sqrt{3(1-\eta)/2}$. If, on the other hand, we detect one more photon
than we inject, then we will remove $n$ photons from the input
mode. This will remove from the output state the subspace that has
fewer than $n$ photons. Thus, if the input state has a component
$\alpha|0\rangle$, and we remove one photon from the mode, this
component will be removed, so that $\alpha$ will be set to zero, and
the information contained in the value of $\alpha$ will be lost. As
far as the operation of a gate is concerned, this results in a failure
that cannot be corrected downstream using feed-forward. However, if we
have already added $n$ photons to the input mode, then we can remove
them without losing any of the coefficients. If we start with an input
state $\alpha' |1\rangle + \beta' |2\rangle + \gamma' |3\rangle$,
inject $k$ photons at the auxiliary port, and remove a photon from
the mode, the resulting transformation for $k>0$ is
\begin{eqnarray}
  \alpha' & = & \alpha \sqrt{(1-\eta)(k+1)/k} \sqrt{\eta}^k  \\  
  \beta'  & = & \beta \sqrt{2(1-\eta)(k+1)/k}  \sqrt{\eta}^{k+1} 
                \left[ 1 - k \xi/2 \right] \\
  \gamma' & = & \gamma  \sqrt{6(1-\eta)(k+1)/k} \sqrt{\eta}^{k+2} \nonumber \\
          &   & \;\;\; \times \left[ 1 - k \xi  + k(k-1) \xi^2 /6 \right] .
  \label{tran3}
\end{eqnarray}
where we have omitted an overall phase factor of $-i$. For $k=0$ the
transformation is $\alpha' = \alpha \sqrt{1-\eta}$, $\beta' = \beta
\sqrt{2\eta(1-\eta)}$ and $\gamma' = \gamma \sqrt{6(1-\eta)}\eta$.

Finally, it is important to note that the transformation that is
performed on the input coefficients depends not only upon the number
of injected and detected photons, but also on the number of photons in
the input mode (that is, on the number of photons which has previously
been added to the input beam). If we both inject and detect $k$
photons, but have an input state with one extra photon
(that is, $|\psi_{\mbox{\scriptsize in}}\rangle = \alpha |1\rangle + \beta |2\rangle
+ \gamma |3\rangle$), then the transformation is
\begin{eqnarray}
  \alpha' & = & \alpha \sqrt{\eta}^{k+1} \left[ 1 - k\xi  \right] \\  
  \beta'  & = & \beta  \sqrt{\eta}^{k+2} \left[ 1 -  2 k \xi + k(k-1)\xi^2 \right] \\ 
  \gamma' & = & \gamma \sqrt{\eta}^{k+3} \left[ 1 - 3 k \xi + 3k(k-1) \xi^2 - \frac{k!}{(k-3)!} \xi^3 \right]  , \nonumber
  \label{tran4}
\end{eqnarray}

Before we consider the use of feed-forward, we investigate whether it
is possible to construct an NS gate with two beam splitters by using
different numbers of input photons than those in the gate described
above. We do this because different options for implementing the gate
might provide alternative options for correcting the output of the
gate using feed-forward.

It will be useful to introduce a compact notation to describe a given
sequence of beam splitters. We will denote a beam splitter with $k$
photons injected into the auxiliary mode, and $n$ photons detected at
the auxiliary output by the pair $(k,n)$. A sequence of two or more
beam splitters will then be denoted $[(k_1,n_1),(k_2,n_2),\ldots]$. In
this notation, the NS gate of Ralph {\em et al.} is $[(1,1),(0,0)]$. Note 
that the configuration $[(n,n),(m,m)]$ is completely equivalent to 
$[(m,m),(n,n)]$, since no photons are added to the beam by either element.
We now explore these configurations for $n,m = 0\ldots 4$. We find that 
$[(1,1),(m,m)]$ can be used to produce an NS gate, however, the success 
probability slowly falls as $m$ is increased. In particular, 
$[(1,1),(1,1)]$ will generate an NS gate with success probability 
$P=0.209$, a little below that of the [(1,1),(0,0)] configuration. 
(This requires the transmitivities $\eta_1=0.2275$ and $\eta_2=0.91968$, 
or vice versa.)
Conversely, configurations of the form $[(n,n),(0,0)]$ will only 
generate an NS gate for $n=1$. Configurations in which both $n$ and $m$ 
are larger than $1$ will generate NS gates but with significantly lower 
success probabilities. We also note that the configuration $[(1,0),(0,1)]$, 
which injects a photon
into the mode at the first beam splitter and removes it at the second
beam splitter, cannot generate an NS gate because all the
transformation factors for the coefficients are positive.  
Thus, as one increases the number of auxiliary photons used, the
success probability of the resulting NS gates get {\em worse} rather
than better, thus the scheme devised by Ralph {\em et al.} and Rudolph
and Pan has what appears to be the highest success probability using 
only two concatenated beam splitters.~\cite{Zou}

We now consider the use of feed-forward to increase the success
probability of a two-element NS gate. To implement this feed-forward
one first measures the auxiliary output of the first beam
splitter. If this output is correct for the implementation of the
gate, then the second beam-splitter is implemented as usual. However,
if the output is not correct, then we route the output beam to a new
set of two beam splitters designed to transform this output so as to
produce the correct transformation for the NS gate.  We expect that
only two beam splitters will be required to do this, since, as
mentioned above, this provides a transformation with two tunable
parameters.

To implement feed-forward we chose a version of the NS gate which has
one photon injected into the first beam splitter. In this case, if we
measure no photons in the auxiliary output, we have an error which is
potentially correctable. As discussed above, an NS gate can be
constructed in this manner by choosing $\eta_1=0.2265$ (using the
configuration [(1,1),(0,0)]) or $\eta_1=0.9197$ or $0.2275$ (using the
configuration [(1,1),(1,1)]). Since the last of these is not
significantly different from the first, we need consider only the
first two.  We will therefore apply various possible correction
circuits to an error with these two values of $\eta_1$ to determine which
provides the highest probability of a successful correction.

\begin{table}
\caption{Success probabilities for three concatenated beam splitters}
\begin{tabular}{|c@{\hspace{0.1cm}}|l@{\hspace{0.1cm}}|l@{\hspace{0.1cm}}
                |l@{\hspace{0.1cm}}|l@{\hspace{0.1cm}}|}
   \hline
   {\bf Sequence} & {$\eta_1$} & {$\eta_2$} & {$\eta_3$} & {$P$}\\ 
   \hline
   (1,0),(0,1),(1,1) & 0.9197 & 0.2947 & 0.2567 & 0.0145 \\
   (1,0),(1,2),(0,0) & 0.9197 & 0.1472 & 0.5137 & 0.0202 \\  
   (1,0),(1,2),(1,1) & 0.9197 & 0.1511 & 0.8398 & 0.0173 \\   
   (1,0),(0,0),(1,2) & 0.9197 & 0.5137 & 0.1472 & 0.0104 \\  
   (1,0),(1,1),(0,1) & 0.9197 & 0.6500 & 0.4182 & 0.0042 \\     
   (1,0),(1,1),(1,2) & 0.2265 & 0.3315 & 0.0531 & 0.0088 \\ 
     ''              & 0.9197 & 0.8690 & 0.1766 & 0.0127 \\  
   \hline
\end{tabular}
\label{tab1}
\end{table}

There are a variety of two-element configurations which can be used
for correction. The only requirement is that the sequence subtract a
photon from the input beam, since the detection of zero photons at the
first beam-splitter has added a photon to the beam. We will examine
sequences which will do this using low numbers of auxiliary
photons. We investigate the sequences [(0,1),(0,0)], [(0,1),(1,1)],
[(1,2),(0,0)] and [(1,2),(1,1)] in which the photon is extracted at
the first beam splitter, and the sequences [(0,0),(0,1)],
[(1,1),(0,1)], [(0,0),(1,2)], and [(1,1),(1,2)] in which the photon is
extracted at the second beam splitter.  To calculate the success
probably for the correction, we consider the sequence of three beam
splitters, where the first is the initial beam splitter at which the
error $(1,0)$ occurs, and the second two are the pair used for
correction. The transmitivity of the initial beam splitter is set at
one of the three possible values given above. We then find all
solutions for the two transmitivities of the second pair which
generate the correct NS gate output, and from these obtain the success
probabilities.  We present the success probabilities for the various
beam splitter triples in table~\ref{tab1}. We have only included 
those that provide a solution with an appreciable success probability. 

Examining the table, note first that only one of the sequences is able
to correct for error when $\eta_1 = 0.2265$. That is, the gate that
uses an auxiliary photon at both inputs has more ways to perform a
correction, and these give higher success probabilities. However, this
gate has itself a reduced success probability over the gate with
$\eta_1 = 0.2265$.  This is similar to the behavior noted by Scheel
{\em et al.}, that if one wants to increase the probability of a
successful correction, then one must reduce the success probability
of the initial gate.

The highest success probability for correction is provided by the
sequence [(1,0),(1,2),(0,0)], and this is only $0.0202$. Our feed-forward 
procedure is therefore only able to increase the success
probability of the initial gate a by small amount. One could consider
further corrections, in which an error at the second beam splitter
could be corrected with a further beam splitter pair. However, such a
correction would involve a sequence of four beam splitters, and we
have seen that moving from two beam splitters to three causes a big
reduction in the success probability. We can therefore expect a sequence
of four beam splitters to produce only a miniscule increase in the
total success probability. It appears therefore that feed-forward,
certainly in the configuration we have considered, is not likely to
provide a means to significantly improve the success probability of
linear optical quantum gates.

{\em Acknowledgments}: This work was supported by The Hearne Institute for 
Theoretical Physics, The National Security Agency, The Army Research Office 
and The Disruptive Technologies Office.


\begin{thebibliography}{99}

\bibitem{KLM}
   E. Knill, R. Laflamme, and G.J. Milburn, Nature {\bf 409}, 46 (2001);
   see also G. G. Lapaire, P. Kok, J. P. Dowling, and J. E. Sipe
   Phys. Rev. A {\bf 68}, 042314 (2003.

\bibitem{Kok06}
  A review of recent developements in linear optical quantum computing 
  may be found in 
  P. Kok, W.J. Munro, K. Nemoto, T.C. Ralph, J. P. Dowling and G.J. Milburn, 
  quant-ph/0512071.

\bibitem{FBRev}
   See, e.g. V.P.Belavkin, Rep. Math. Phys. {\bf 45}, 353 (1999); 
   A. C. Doherty, S. Habib, K. Jacobs, H. Mabuchi, S. M. Tan, Phys. Rev. A {\bf 62}, 
   012105 (2000). 
   A recent introductory review of feedback in quantum systems is given in
   K. Jacobs, Proceedings of the 6th Asian Control Conference (in press), 
   Eprint: quant-ph/0605015.

\bibitem{Scheel04} 
  S. Scheel and N. L\"utkenhaus, New J. Phys. {\bf 6}, 51 (2004).

\bibitem{Eisert05}
  J. Eisert, Phys. Rev. Lett. {\bf 95}, 040502 (2005).

\bibitem{Knill03}
  E. Knill, Phys. Rev. A {\bf 68}, 064303 (2003).
  
\bibitem{Scheel05} 
   S. Scheel, W.J. Munro, J. Eisert, K. Nemoto, P. Kok, Phys. Rev. A {\bf 73}, 034301 (2006).

\bibitem{Ralph02}
  T. C. Ralph, A. G.White, W. J.Munro, and G. J. Milburn, Phys. Rev. A, {\bf 65} 012314 (2002).

\bibitem{Rudolph01}
  T. Rudolph and J. W. Pan, quantph/0108056.

\bibitem{Zou}
  While we are concerned here only with gates in which the beam splitters 
  act separately in a concatenated sequence, it is possible to construct 
  a gate with two auxiliary photons and two beam splitters 
  in which this is not the case. Such a gate is given in  
  X. Zou, K. Pahlke, and W. Mathis, Phys. Rev. A {\bf 65}, 064305 (2002).
                                
\end{thebibliography}
\end{document}